\documentclass[12pt]{article}

\usepackage{graphicx} 
\usepackage{amsmath} 
\usepackage{amssymb}  
\usepackage{amsthm}
\usepackage{physics}

\title{Model Free Control =  Youla-Kucera PI Control}

\author{Ioan~Dor\'{e}~Landau\\ \small Univ. Grenoble Alpes, CNRS, Grenoble INP, GIPSA-lab, 38000 Grenoble, France}

\begin{document}
\maketitle


\section*{Abstract}
This note shows that Model Free Control strategy is simply a Youla-Kucera parametrized PI control strategy which can be also re-formulated as a standard PI control both in discrete-time and in continuous time. The note shows also the influence of the estimated values of the plant gain upon the performance of the control system.

\section*{Keywords}
 PI control, Model free control, Youla-Kucera observer, continuous time, discrete-time



\section*{List of Acronyms}\label{Acronyms}
\noindent
MFC - Model Free Control\\
PI - Proportional + Integral Controller\\
YK - Youla Kucera\\
\section{MFC - original formulation}
In MFC \cite{Fliess13} it is assumed that the output of the plant to be controlled an be expressed as:
\begin{equation}
y'(t)=F(t)+ \alpha u(t)
\label{eq:1}
\end{equation}
where $u(t) $is the plant input, $y(t)$ is the output and F(t) is a disturbance. 
In fact, it is assumed that the output of the plant can be represented by an integrator with the gain $\alpha$ subject to an input disturbance $F(t)$. 
In MFC papers (\cite{Fliess13},\cite{Robu22}) $F(t)$ is termed "ultra local model" of the plant. However $F(t)$ is not a model of the plant because the dependence on u(t) is not taken into account and in fact it is implicitely assumed that the plant can be represented by \eqref{eq:1}, i.e. it is assumed that all the plant can be represented by an integrator with a gain and subject to a disturbance.  
In MFC, F(t) is theoretically estimated through
\begin{equation}
\hat{F}(t)=y'(t) - \alpha u(t)
\label{F1}
\end{equation}
and the theoretical control law is:
\begin{equation}
	\label{eq_u}
	u(t) =-\left\lbrace \hat{F}(t)-y^{*'}(t)+ K_p(y(t)-y^*(t))\right\rbrace g
\end{equation}
where $y^*$ is the reference and $g=1/\alpha$.\\
Why it is relatively easy to get an order of magnitude of $\alpha$, getting a very precise estimation is difficult without using plant model identification. In practice the value of  the true $\alpha$ should be replaced by its estimation $\hat{\alpha}$ and $g$ by its estimation $\hat{g}=1/\hat{\alpha}$. \\
The main problem for the implementation of this control law is the fact that in practice $y'(t)$ can not be exactly obtained and therefore it should be approximated. There are several options: time discretization (see Section \ref{DT}) or a continuous time framework implementation using "state variable filters"(see Section \ref{CTI}).\\
The analysis which follows can be applied straightforwardly also to the other control laws considered in \cite{Fliess13},\cite{Lafont14}.\\
A related work is \cite{AndreaNovel10}.
\section{Discrete time case}\label{DT}
Unfortunately in practice $y'(t)$ can not be exactly obtained and one option is to use time discretization.
 In addition the gain $\alpha$ is not exactly known and should be replaced by its estimation  $\hat{\alpha}$. Accordingly the inverse of the plant gain denoted $g$ will be replaced by $\hat{g}=1/\hat{\alpha}$.\\

\begin{figure}[htb!]
	\centering
	\includegraphics[width=0.9\columnwidth]{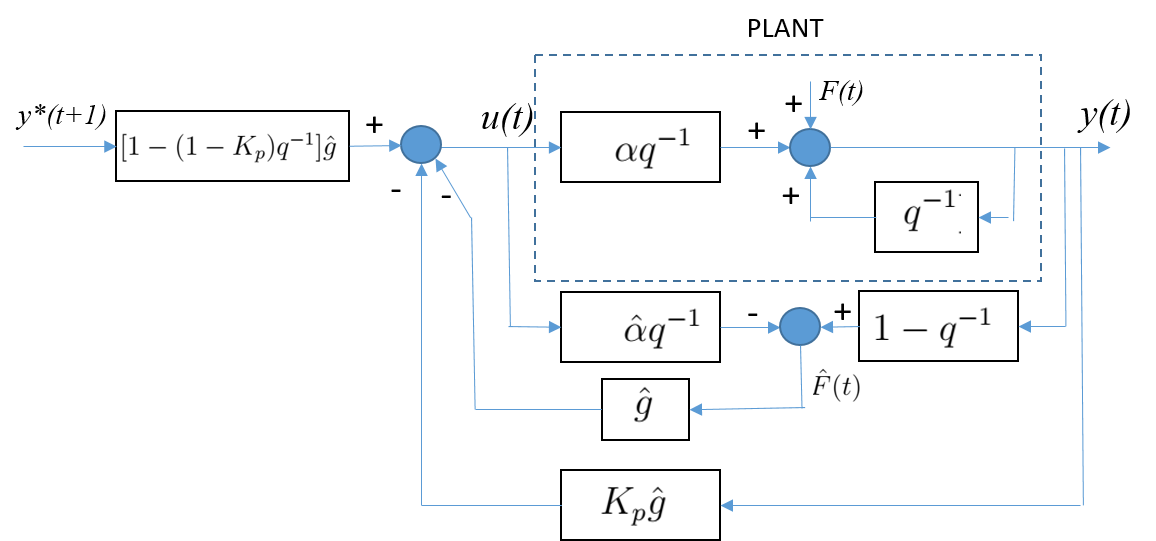}
	\caption{Model Free Control Diagram }
	\label{fig:MFC1}
\end{figure}

The plant equation becomes in discrete time \cite{Robu22} :
\begin{equation}
y(t+1)=y(t)+F(t+1)+\alpha u(t)
\end{equation}
where F in discrete time is the continuous time F multiplied by $T_s$ ( $=F T_s$) where $T_s$ is the sampling period. 
Eq. \eqref{F1} becomes in discrete time:
\begin{equation}
\hat{F}(t)= y(t)-y(t-1)-\hat{\alpha}u(t-1) = (1-q^{-1})y(t)-\hat{\alpha}q^{-1}u(t)
\label{FD}
\end{equation}
This is just a Youla - Kucera disturbance observer used in many control schemes and in particular in adapive regulation (see \cite{Valentinotti03}, \cite{Landau05a}).
The control law given in Eq. \eqref{eq_u} becomes in discrete time:
\begin{equation}
u(t)=-[\hat{F}(t)-(y^*(t+1)-y^*(t))]\hat{g}-K_p[y(t)-y^*(t)]\hat{g}
\label{ud}
\end{equation}
where $y^*$ is the reference which is known one step ahead (we know at time $t$ the value which we want to attain at $t+1$) and $\hat{g}=\frac{1}{\hat{\alpha}}$. Combining Eq. \eqref{FD} with Eq. \eqref{ud}  the diagram of the control scheme shown in Fig. \ref{fig:MFC1} is obtained.

\begin{figure}[htb!]
	\centering
	\includegraphics[width=0.9\columnwidth]{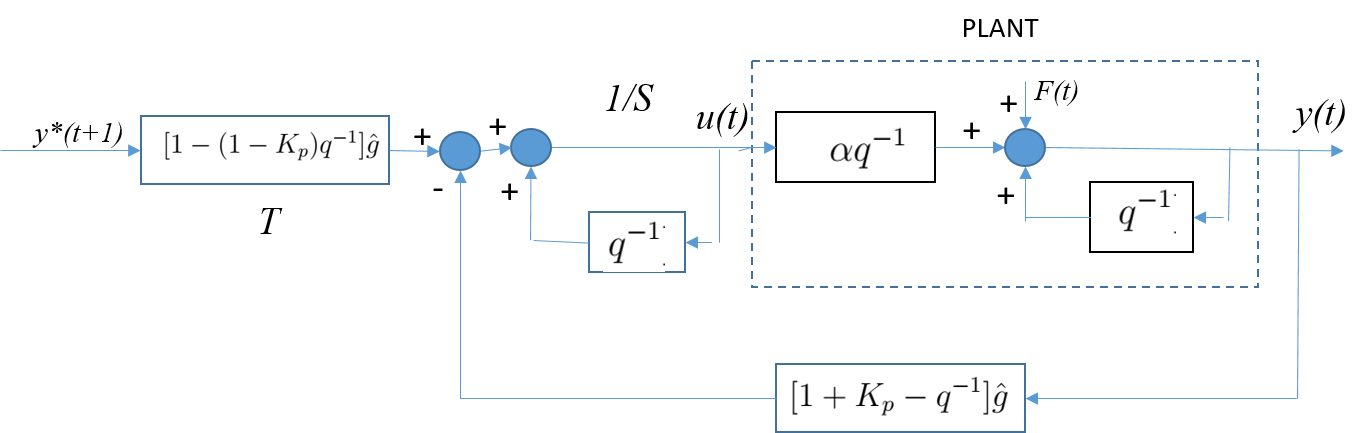}
	\caption{Model Free Control Diagram - R-S-T form.}
	\label{MFC2}
\end{figure}

Taking into account Fig.\ref{fig:MFC1}, Eq.\eqref{ud} can be rewritten as: 
\begin{equation}
S(q^{-1})u(t)=T(q^{-1})y^*(t+1)-R(q^{-1})y(t)
\end{equation}
where:
\begin{equation}
S(q^{-1})=1-q^{-1}=1+s_1q^{-1}
\end{equation}
\begin{equation}
T(q^{-1})=[1-(1-K_p)q^{-1}]\hat{g}=t_0+ t_1q^{-1}
\end{equation}
\begin{equation}
R(q^{-1})=[1+K_p-q^{-1}]\hat{g}=r_0+r_1q^{-1}
\end{equation}
The controller is a digital PI controller in R-S-T form (see \cite{Landau05b})
The corresponding equivalent diagram is shown in Fig.\ref{MFC2}.\\
The poles of the closed loop are given by:
\begin{multline}
P=AS+BR=(1-q^{-1})(1-q^{-1})+\alpha q^{-1}[1+K_p-q^{-1}]\hat{g}\\
=1-[2-\hat{g}\alpha(1+K_p)]q^{-1}
+(1-\hat{g}\alpha)q^{-2}
\label{P}
\end{multline}
where $A=1-q^{-1}$ , $B=\alpha q^{-1}$ and the transfer operator from $y*$ to $y$ is given by:
\begin{equation}
S_{yy*}=\frac{TB}{P}= \frac{[1-(1-K_p)q^{-1}]\hat{g}\alpha q^{-1}}{1-[2-\hat{g}\alpha(1+K_p)]q^{-1}
+(1-\hat{g}\alpha)q^{-2}}
\end{equation}
The I/O steady state gain is given by
\begin{equation}
S_{yy*}(1)= \frac{K_P\hat{g}\alpha}{\hat{g}\alpha K_p)}=1
\end{equation}
Of course it is possible to obtain a unit gain also for $\hat{\alpha} \neq\alpha$ by taking $T(q^{-1})=R(q^{-1})$ or $T=R(1)$.\\

In the case $\hat{g}=\frac{1}{\alpha}$ the poles of the closed lop are given by:
\begin{equation}
P=1-(1-K_p)q^{-1}~;~~for~~\hat{g}=1/\alpha
\end{equation}
Stability of the closed loop is assured for $0<K_p<2$ (or $ 0<K_p<1$ if one wants a pole with an a periodic response)  while this is not true when $\hat{\alpha} \neq\alpha$. In this later case $K_p$ has to be chosen such the $P$  given in Eq.\eqref{P} has its roots inside the unit circle (one can use well known stability criteria). It is also possible solving a Bezout equation to find the values of the coefficients of $R(q^{-1})$ for given desired closed loop poles \cite{Landau05b}.\\
In \cite{AndreaNovel10} a comparison of classical PID configurations and MFC PID configurations is provided by discretization of the continuous time formulation of both structures. However, it is our opinion that the Youla-Kucera formulation allows to give a common framework for the two types of formulation. 
\section{Continuous time implementation}\label{CTI}
\subsection{Use of state variable filters}
\begin{figure}[htb!]
	\centering
	\includegraphics[width=0.9\columnwidth]{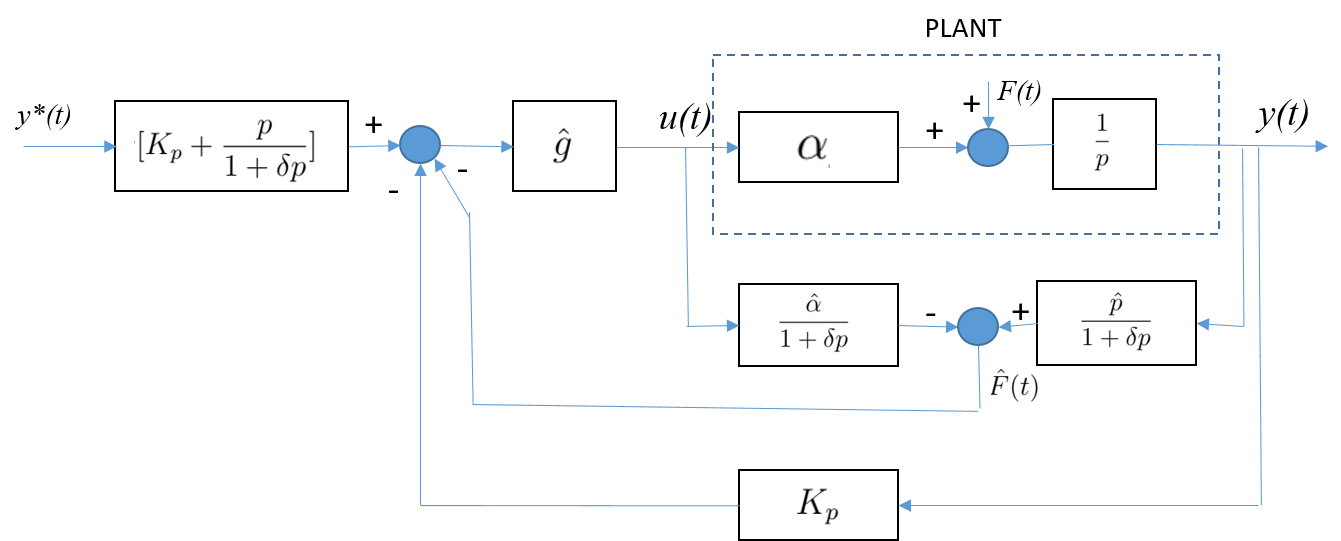}
	\caption{Model Free Control Diagram - Continuous time.}
	\label{MFC_c1}
\end{figure}

 In practice $y'(t)$ and $y^{*'}(t)$ should be replaced by approximations and the true $\alpha$ should be replaced by its estimation $\hat{\alpha}$ and $g$ by its estimation $\hat{g}=1/\hat{\alpha}$.\\
To approximate the derivatives a first solutions is to use a first order "state filter" generating filtered variables allowing to compute their derivatives. Specifically one has:
\begin{equation}
y_{f}'(t)=-\frac{1}{\delta}y_f(t)+\frac{1}{\delta}y(t)
\end{equation} 
where $\delta$ is called filtering time constant.\\
In operator form we will have
\begin{equation}
y_f(t)=\frac{y(t)}{1+\delta p}
\end{equation} 
\begin{equation}
u_f(t)=\frac{u(t)}{1+\delta p}
\end{equation} 
\begin{equation}
y^*_f(t)=\frac{y^*(t)}{1+\delta p}
\end{equation} 
The estimation of F(t) denoted $\hat{F}(t)$ will be given by:
\begin{equation}
\hat{F}(t)=\frac{p}{1+\delta p}y(t)- \frac{\hat{\alpha}}{1+\delta p}u(t)
\label{eq:fc}
\end{equation} 
This is nothing else that a particular Youla-Kucera disturbance observer\footnote{Indeed this corresponds to the factorization of the plant as the product G=$ND^{-1}$ of stable transfer functions used in Youla-Kucera parametrization. See
 \cite{Anderson98}}.\\
The control law becomes:
\begin{equation}
u(t) =-[\hat{F}(t)-y^{*'}_{f}-K_p(y(t)-y^*(t))]\hat{g}
\label{eq:ucf}
\end{equation}
This leads to the control scheme shown in Fig.\ref{MFC_c1}
Introducing Eq. \eqref{eq:fc} in Eq. \eqref{eq:ucf}, one gets (see also Fig \ref{MFC_c1}):
\begin{equation}
u(t)=\hat{g}[K_p +\frac{p}{1+\delta p}]y^*(t)-\hat{g}[K_p+\frac{p}{1+\delta p}]y(t)- \frac{1}{1+\delta p}u(t) 
\end{equation}
This equation can be-rewritten (after grouping the terms containing $u(t)$) as:
\begin{equation}
u(t)=\frac{(\hat{g}/\delta)[K_p+(1+K_p \delta)p]}{p}(y^*(t)-y(t))
\end{equation}
 which is nothing else that a PI controller in operator form. Using a parallel implementation one gets the $P_{PI}$ and (proportional action) and $I_{PI}$(integral action) as:
 \begin{equation}
u(t)=(P_{PI}+\frac{I_{PI}}{p})(y^*(t)-y(t));~ P_{PI}=(\hat{g}/\delta)(1+K_p \delta);~ I_{PI}=(\hat{g}/\delta)K_p
 \end{equation}
 The transfer operator from the reference to the output is given by:
 \begin{equation}
 S_{yy^*}=\frac{\alpha\hat{g}[K_p+(1+\delta K_p)p]}{\hat{g}\alpha K_p+ \hat{g}\alpha(1+\delta K_p)p+ \delta p^2}
 \end{equation}
 It can be verified that the steady state gain is 1 and that the stability of the system will depend upon  $K_p$, $\hat{g}$ and $\delta$.\\
 The analysis is similar when higher  order state variable filters are used. \\
\section{Comments}
\begin{itemize}
\item Despite that the authors of \cite{Fliess13} and of subsequent papers like \cite{Robu22} and  \cite{Lafont14} claim that they do not use a model of the plant, indeed they do. They assume that the input-output behavior of the plant is described by the model of Eq. \eqref{eq:1} and this model is used for the design of the controller.
\item This model requires the knowledge of the plant gain and therefore an estimation procedure has to be considered (partial system identification).
\item The essence of the controller proposed is to use a Youla -Kucera observer for estimating the disturbance term $F(t)$ in Eq. \eqref{eq:1} leading to a re-parametrized PI(PID) controller. It is surprising that the authors were not aware about Youla-Kucera parametrization and the use of this type of observers in the control literature and applications.
\item There are many applications which can be solved using   PI controllers. The fact that the authors claims  a significant number of applications, it just simply means that the applications considered enter in the category of control  applications which can be solved with a lower order controller. No applications are reported (as far as  my knowledge goes) on systems featuring multiple low damped oscillatory modes.
\item The main issue in applying PI controllers is the availability of a performance oriented tuning procedure taking into account tracking and regulation specifications. Unfortunately the authors of \cite{Fliess13} and subsequent papers do not propose any systematic tuning procedure except a trial and error approach. The fact that it may exists a couple of parameters providing good performance which has to be found by trial and error is not the answer. A good tuning procedure should provide directly the values of the controller parameters from the performance specifications and information acquired from the system. Such procedures exits. See for example: \cite{Astrom88b}, \cite{Voda95} among many other references.
\item The authors do not propose any robustness analysis with respect to model discrepancies (except some simulations). However once their controller is expressed in standard form either as  a two degree of freedom controller or simply as a filter in operator form acting on the control error, available results can be used for assessing the robustness.

\item The term "intelligent" used by the authors is in my opinion inappropriate. There is no any added intelligence in replacing the standard PI formulation by  a re-parametrization using a Youla-Kucera observer. The term"intelligent" can eventually associated to controllers including some automatic data-driven tuning procedure.

\end{itemize}
%
 
\bibliographystyle{abbrv}
\bibliography{Bibliography}

\end{document}